\documentstyle[12pt]{article}

\newcommand{\Section}[1]{\section{#1}\setcounter{equation}{0}}

\def\sqr#1#2#3#4{{\vcenter{\vskip -#3 pt\hbox{\kern #4 pt\vbox{\hrule height
.#2 pt\hbox{\vrule width .#2 pt height #1 pt\kern #1 pt\vrule width .#2 pt}
\hrule height .#2 pt}\kern #4 pt}}}}

\def\QED{\hbox{\kern 1pt\vrule width 3pt height 7pt}}
\def\d{\delta}

\def\be{\begin{equation}}
\def\bea{\begin{eqnarray}}

\def\ee{\end{equation}}
\def\eea{\end{eqnarray}}

\def\C{\rm {I\kern-.520em C}}
\def\w{\rm {WZNW }}

\begin{document}
\begin{titlepage}
\begin{center}
\vskip .2in
\hfill
\vbox{
    \halign{#\hfil         \cr
            hep-th/9410158 \cr
            October 1994   \cr
           } 
      }  
\vskip 1cm
{\large \bf Vector-Chiral Equivalence in Null Gauged WZNW Theory }
{\large \bf }
\vskip 0.5in

{\bf Farhad Ardalan}
\vskip .1 in
\center{and}
\vskip .1 in

{\bf Amir Masoud Ghezelbash}
\vskip .25in
{\em  Institute for Studies in Theoretical Physics and Mathematics, \\
 P.O. Box 19395-5746, Tehran, Iran. \\
Department of Physics, Sharif University of Technology, \\
P.O. Box 11365-9161, Tehran, Iran. }
 \vskip .5in

\end{center}
\begin{abstract}
We consider the standard vector and chiral gauged \w models by their gauged
maximal null subgroups and show that they can be
mapped
to each other by a special transformation.
We give an explicit expression for the map in the case of
the classical Lie groups $ A_N $, $ B_N $, $ C_N $, $ D_N $, and note its
connection with the duality map for the Riemmanian globally symmetric spaces.
\end{abstract}

\end{titlepage}
\newpage
\def \dbar {\bar \partial}
\def \ar {\rightarrow}
\def \d  {\partial}

\Section{Introduction}

Duality as a abelian or non-abelian symmetry of conformal field theories has
been studied
extensively in the last few years \cite {SFETSOS3},\cite{ALVAREZ},\cite{BUSH}.
In \w models, this symmetry, when originating from an abelian subgroup
of the isometry of the conformal background, has been better
understood.

In fact, it has been shown that automorphisms of the group $ G $ of the model
are
responsible for their duality transformations \cite{OBERS}. Specifically, it
has been
proved that the target manifold of the \w model with the group $ G $, obtained
by
vector gauging its abelian subgroup $ H $, is dual to the manifold obtained by
axially gauging the same subgroup \cite{KIRITSIS}; and this dual transformation
is
implemented by an automorphism of $ H $, on the left action of $ H $ on $ G
$.\newline
The above considerations are generally valid, whether the subgroup $ H $ is
semisimple or not. But, it has been known that when $ H $ is not
semisimple, certain new phenomena could occur. In particular when $ H $
is nilpotent,\footnote{To be exact one should consider null groups,
which are
slightly more general than the nilpotent groups.} the target space shows
unusual
features and its dimension reduces unexpectly \cite{ARDALAN},
\cite{ARDALAN2},\cite{KUMAR}.

In other words, it was shown that
in the
usual vector gauging
of $ SL(2,R) $ \w model, by its nilpotent one dimensional subgroup
, the target space metric becomes one
dimensional and the effective action of the gauged model reduces to that of
the Liouville
field theory.
Similar behaviour for $ SO(3,1) $ vector gauged \w, by its two
dimensional Euclidean subgroup,
is also observed \cite{GHEZEL}. In this case, although we expect a three
dimensional target space, the resultant target space is only one
dimensional.

It has been suspected that the reduction of the degrees of freedom of the
target manifold for those coset $ G/H $ \w models, when a nilpotent subgroup $
H
$ is vectorially gauged, is related to the obvious dimensional reduction of the
corresponding chiral gauged model when the left and right actions are
independent and in two different subgroups isomorphic to $ H
$ \cite{FEHER}.

This dimensional reduction occures for the Toda theories by gauging
the \w model, with the total
left and
right null subgroups $ G_+ $ and $ G_- $ generated by positive and negative
step operators in the Cartan-Weyl decomposition of Lie group $ G $.
On the other hand, if the dimension of gauged
subgroups are less than the dimension of $ G_+ $ and $ G_- $, then in the
resulting $\sigma - $model, one obtains the
Toda theory with an interaction term that comes from the remaining
part of ungauged directions of $ G_+ $ and $ G_- $ \cite{KLIMCIK}.
The Toda structure is also obtained, however, by simply gauging the
nilpotent subgroup vectorially, in the case of $ SO(3,1) $
\cite{GHEZEL}.

In this letter we will construct a one to one map from the vector gauged model
to the chiral gauged model $ G/H $, when $ H $ is nilpotent, explaining the
dimensional reduction of the vector gauged model and find that this map is in
fact an automorphism of $ G $. This automorphism turns out to be the dual
map of the well known Riemannian globally symmetric manifolds.

In section two, we will briefly recall the construction of the gauged \w
models, and describe the reduction of the dimension of the target manifold for
$ SL(2,R) $ by its nilpotent subgroup when gauging it vectorially. In
section three we construct the
automorphism relating the vector gauged \w model to the chiral case,
and generalize this automorphism to the case of other simple Lie groups
gauged by their maximal null subgroups.
In section four, we
note on the relation of this map to the corresponding duality map
for Riemannian globally symmetric manifolds.

\Section{Structure of gauged WZNW models }

Let us recall first the structure of vector and chiral gauged \w. The $ G/H $
vector gauged \w action \cite{WITTEN}, \cite{GAWEDZKI}
\bea S_V(g,{\bf {A}},{\bf {\bar A}}
)=S(g)+{k\over 2\pi}\,\int
\,d^2z\,
Tr(-\,{\bf {\bar A}} g^{-1} \d g+{\bf {A}}\dbar {g} g^{-1}-{\bf {A}}{\bf {\bar
A}}+g{\bf {\bar A}}g^{-1}{\bf {A}}\,), \eea $$
S(g)={k\over 4\pi}\,\int
\,d^2z\,
Tr(\,g^{-1}\partial gg^{-1}\partial
g\,)-{k\over 12\pi}\,\int \,
Tr(\,g^{-1}dg\,)^3,
$$
is invariant under the gauge transformations
\be g \rightarrow h^{-1}g\,h,{\bf {A}}\rightarrow h^{-1}\,({\bf
{A}}-\partial)\,h,{\bf {\bar A}}\rightarrow h^{-1}\,({\bf {\bar A}}-\bar
\partial)\,h,
\ee where $ h=h(z,\bar z) $ is a group element in subgroup $ H $ and
$ {\bf {A}}$ , $ {\bf {\bar A}} $ take their values in the algebra $ \cal{L}
$$(H)
$ of the subgroup $ H $. Parametrising
$ {\bf {A}}$ and $ {\bf {\bar A}} $ in terms of the group elements $ l $, $
\bar
l $ of $ H $  $$ {\bf {A}}=\d l \, l^{-1} ,\quad {\bf {\bar A}}=\dbar \bar l \,
 \bar l^{-1}, $$ one can write the above form of vector action (2.1), using the
Polyakov-Wiegmann identity \cite{POLYAKOV},
as the difference of two terms \cite{SFETSOS}
\be S_V(g,{\bf {A}},{\bf {\bar A}})=S(l^{-1} g \bar l)-S(l^{-1} \bar l). \ee
Each term is invariant under gauge transformations,
$ g \ar h^{-1} g h , l \rightarrow h^{-1} l , \bar l \rightarrow h^{-1} \bar l
$. On the other hand, the chiral \w action \cite{CHUNG}, \cite{SFETSOS2},
\cite{SFETSOS4}
\bea S_C(g,{\bf {A}},{\bf {\bar A}})=S(g)+{k\over 2\pi}\,\int
\,d^2z\,
Tr(-\,{\bf {\bar A}} g^{-1} \d g+{\bf {A}}\dbar {g} g^{-1}+g{\bf {\bar
A}}g^{-1}{\bf {A}}\,), \eea
is invariant under the following transformations
\be  g \rightarrow h^{-1}g\,\bar h,{\bf {A}}\rightarrow
h^{-1}\,({\bf {A}}-\partial)\,h,{\bf {\bar A}}
\rightarrow \bar h^{-1}\,({\bf {\bar A}}-\bar \partial)\,\bar h, \ee
where $ h=h(z) $ belongs to the subgroup $ H_1 $, and $ \bar h=\bar h(\bar z) $
belongs to another subgroup $ H_2 $ of $ G $. $ {\bf {A}}$ takes its value in
$\cal {L}$
$(H_1)$ and $ {\bf {\bar A}} $ in $ \cal {L}$ $(H_2)$. Parametrising $ {\bf
{A}}, {\bf {\bar A}} $ in
terms of $ l \in H_1 $ and $ \bar l \in H_2 $ as before, $ S_C(g,{\bf {A}},{\bf
{\bar A}}) $
can be written as
$$ S_C(g,{\bf {A}},{\bf {\bar A}})=S(l^{-1} g \bar l)-S(l^{-1})-S(\bar l). $$
We see that these chiral transformations don't fix and elliminate any
dynamical degree of
freedom of $ g $ , since $ h $ and $ \bar h $ are only holomorphic and
antiholomorphic
functions respectively and not arbitrary functions of $ z $ and $ \bar z $, in
contrast to vector case.
But if we restrict $ H_1 $
and
$ H_2 $ to be null or nilpotent subgroups of $ G $ (A group $ H $ is null if
$Tr(N_i^2
)=0, $ for every $ i $ where $ N_i $'s are
generators of algebra $\cal {L}$ $
(H) $; and nilpotent if $ N_i^2=0 $)
then $ S(l^{-1})=S(\bar l)=0 $ and $ S_C(g,{\bf {A}},{\bf {\bar A}})=S(l^{-1}
g \bar
l) $ is invariant under truly gauge transformations (2.5) with $
h=h(z,\bar z) \in H_1 $
and $ \bar h=\bar h(z,\bar z) \in H_2 $ or $ g \ar h^{-1} g \bar h, l \ar
h^{-1} l,\bar l \ar \bar h^{-1} \bar l $. In the following
we choose $ H_1=G_+ $ (in the vector case we also choose $ H=G_+ $ ) and $
H_2=G_- $ respectively,
where $ G_+ $  $ (G_-) $ is the null subgroup of $ G $ generated by the set of
positive (negative) step operators.

Now,
let us consider $ SL(2,R)/E(1) $ vector gauged \w model. The $ E(1) $
subgroup is generated by the nilpotent positive step operator $ \pmatrix {0 &
1 \cr 0 & 0} $ in the triangular decomposition of the Lie algebra $ sl(2,R) $.
The
gauge part of the vector gauged action $ S_V(g,{\bf {A}},{\bf {\bar A}}) $
with the parametrization
of $ g=\pmatrix {a & u \cr -v & b } $,$ ab+uv=1 $, is
\be {-k \over \pi} \int d^2z \{ \bar A(v \d a-a \d v)+A(b \dbar v-v \dbar
b)+v^2 A \bar A \}, \ee
while the gauge part of the chiral gauged action $ S_V(g,{\bf {A}},{\bf {\bar
A}}) $,when $ G_+ $ and $ G_-
$ are generated by $ \pmatrix {0 & 1 \cr 0&0} $ and $ \pmatrix {0 & 0 \cr
1 & 0}, $ is
\be {-k \over \pi} \int d^2z \{ \bar A(b \d u-u \d b)+A(b \dbar v-v \dbar
b)-b^2 A \bar A \}. \ee
The common part $ S(g) $ of the action for both cases is
$$ {k \over {2 \pi }} \int d^2z \{ b^2 \d a \dbar a+b u( \d a \dbar v+\d v
\dbar a)+u^2 \d v\dbar v+b v (\d u \dbar a+\d a \dbar u) $$ $$ -a b (\d u \dbar
v+\d
v \dbar u) -u v (\d b \dbar a+\d a \dbar b)+a u (\d b \dbar v+\d v
\dbar v)+v^2
\d u \dbar u $$ $$ +a v (\d u \dbar b+\d b \dbar u)+a^2 \d b \dbar b+\int
{{d\,a}\over a} (\d u \dbar v-\d v \dbar u) \}. $$
The term involving $ \ln a $ comes from the Wess-Zumino term and can be
written in the equivalent form $ \ln u (\d a \dbar b-\d b \dbar a) $
\cite{WITTEN2}.
The gauge symmetry (2.2) forces
us to choose $ a=0 $ or $ b=0 $, in any case upon integration of the action
over
gauge fields $ A, \bar A $, the resultant effective action leads to a target
space
metric and a dilaton field that are proportional to $ d \phi^2 $ and $ \phi
$ respectively, where $ \phi=\ln v $ (The antisymmetric tensor is zero in this
gauge). Obviously one dynamical degree of freedom has evaporated. A similar
feature
is observed for $ SO(3,1) $ when vector gauging it by its three dimensional
Euclidean group \cite{GHEZEL}.
In this case two generators of $ E(2) $ group that correspond to translation in
two
dimensional plane are null and the resultant effective action and dilaton
field
are one dimensional.
\Section{Equivalence of vetor and chiral gauging}

In the last section we saw that in vector gauging a \w model by a
subgroup while is null (or when a part of the subgroup
is null) a dimensional reduction in the effective action of the theory
occured. To explain this reduction,
we consider the $ SL(2,R) $ case first.
It is not hard to see that the following map changes the vector gauged action
(2.6) to the chiral gauged action (2.7)
$$ a \ar -u, u \ar a, v \ar b, b \ar -v, $$ \be A \ar A,
\bar A \ar - \bar A, \ee
or in the compact form $ g \ar g \tau_2,{\bf {A}}\ar {\bf {A}},{\bf
{\bar A}} \ar \tau_2^{-1} {\bf {\bar A}} \tau_2
$, where \be \tau_2=\pmatrix {0 & 1 \cr -1 & 0 }. \ee Note that the
$ S(g) $ part is invariant under this map.
Clearly,from the form of the action $ g \ar g \tau_2 $ one can see that $
\tau_2 $ is a symmetry of the vector gauged theory. On the other hand, $\tau_2
$ maps the vector gauged theory onto the chiral case, then establishing the
equivalence of the two theories. But, in the case of the chiral gauge theory
there is only one degree of freedom. In fact, from chiral symmetry (2.4), one
can set $ u=v=0 $. This, therefore explains the evaporation of the degree of
freedom in the target space metric and dilaton of the vector gauged theory
alluded to above \cite{ARDALAN},\cite{ARDALAN2}.
Note also that the target space metric is the free field part
of the
Toda-like action with its coordinates, the parameters of the Cartan
part of the Gauss decomposition of the group element $ g $; and a linear sum of
these parameters is the dilaton field \cite{KLIMCIK}
{}.

In the part of this article we will attempt to generalize the above results
to higher dimensional groups. We may repeat the same calculation for $
SL(3,R) $ gauged \w model, $ G_+
$ subgroup is generated by the three positive triangular matrices $
e_{ij}; j
< i ; i,j=1,2,3 $ with $ e_{ij} $ the matrix with one at the $
(i,j) $
entry and zero everywhere; similarly, the $ G_- $ subgroup is generated by
$ e_{ij}; j > i ; i,j=1,
2,3 $. We use the matrix representation of $ g $ of the form

\be \pmatrix {a & b & c  \cr d & e & f \cr h & i & j
 } .\ee
Then a simple calculation shows that the
vector gauged $ SL(3,R)/G_+ $ \w is converted to the chiral gauged $ SL(3,R) $
with the following transformation on the $ g, {\bf {\bar A}} $
$$
a \ar c,b \ar b,c \ar -a,d \ar f,e \ar e,f \ar -d,h \ar j $$ $$
i\ar i,j \ar -h
$$
$$
\bar A_1 \ar \bar A_3,\bar A_2 \ar - \bar A_2,\bar A_3 \ar - \bar A_1
$$
or in the compact form as $ g \ar g \tau_3, {\bf {A}}\ar {\bf {A}}, {\bf {\bar
A}} \ar \tau_3^{-1}
{\bf {\bar A}} \tau_3 $, where \be \tau_3=\pmatrix {0 & 0 & -1 \cr 0 & 1
& 0 \cr 1 & 0 & 0 }. \ee
The vector gauged \w action is obviously invariant
under these transformations and we therefore get the equivalence of the
vector and chiral gauged theories.
Generally, a similar result exists for $ SL(2 N,R) $.
In other words, under the transformation \be g \ar g \tau_{2
N},
{\bf {A}}\ar {\bf {A}},{\bf {\bar A}} \ar \tau_{2 N}^{-1} {\bf {\bar A}}
\tau_{2 N}, \ee the $ SL(2 N,R) $ vector
gauged \w model converted to the chiral \w model, where
$ \tau_{2 N}  $ is an off-diagonal matrix with $ N $ minus ones and $ N $
plus ones,
$ \tau_{2 N}= \sum_{i=1}^{N}
e_{i,2N+1-i}-\sum_{i=N+1}^{2 N} e_{i,2 N+1-i}. $ Similarly for
$ SL(2N+1,R) $ ,
the map
is
\be
\tau_{2 N+1}=- \sum_{i=1}^{N} e_{i,2N+2-i}+\sum_{i=N+1}^{2 N+1} e_{i,2 N+2-i}.
\ee
Next, we consider the gauged \w actions with $ SO(2N,R) $ group manifold.
The first nontrivial
case is $ SO(4,R) $; the $ G_+ $ subgroup is generated
by $ e_{14}-
e_{23}, e_{12}-e_{43} $, and $ G_- $ by $ e_{32}-e_{41}, e_{21}-e_{34} $
\cite{BAU}, and the equivalence map is $ s_{4} $, with
\be g \ar gs_4 , {\bf {A}}\ar {\bf {A}}, {\bf {\bar A}} \ar s_4^{-1} {\bf {\bar
A}}
s_4, \ee where $ s_4=\pmatrix {0 & 0 & 1 & 0 \cr
                0 & 0 & 0 & 1 \cr
                -1 & 0 & 0 & 0 \cr
                 0 & -1 & 0 & 0 }    $.
A similar but
long
calculation shows the same equivalence between $ SO(2N,R) $ vector
and chiral gauged \w theories.
In this case the transformations are like $ SO(4,R) $ case (3.8), with $ s_{4}
$
replaced by $ s_{2 N} $ \be g \ar g s_{2 N},{\bf {A}}\ar {\bf {A}},{\bf {\bar
A}} \ar s_{2 N}^{-1}
{\bf {\bar A}} s_{2 N}, \ee where
\be s_{2N}=\sum_{i=1}^{N}
e_{i,i+N}-\sum_{i=N+1}^{2 N} e_{i,i-N}. \ee
In a similar way the following operator
$ s_{2N+1} $ relates the vector and
chiral gauged theories with the group $ SO(2N+1) $, where
\be
s_{2N+1}=(-)^N \{ e_{11} - (\sum_{i=2}^{N+1} e_{i,N+i}+\sum_{i=N+2}^{2N+1}
e_{i,i-N}) \}.
\ee
At last, let us consider the case of
$ SP(2N,R) $ gauged \w actions, the simplest being
$ SP(2,R) $. Here $ G_+ $ and $ G_- $ subgroups
are exactly the same as in the $ SL(2,R) $ case and the $ \tau_2 $ matrix that
of equation (3.2) relates the
vector gauged theory to the chiral case.
Similar
results exist for $ SP(2N,R) $ just as the transformations
(3.9) and (3.10) for the $ SO(2N,R) $ case.\newline
\vskip .2in
\Section{Concluding Remarks}

In the preceding sections, with the help of the operators $
\tau_{2N}, \tau_{2N+1}, s_{2N},s_{2N+1} $, we established the equivalence of
the vector gauged action with that of the chiral gauged case.
In fact, we learn from this equivalence
that, when vector gauging \w by its null subgroup, in addition to the
apparent
symmetry (2.2), there is in another symmetry that is induced
from the chiral gauged version and is as follows
\be g \rightarrow h'g\,h,{\bf {A}}\rightarrow h'\,({\bf
{A}}-\partial)\,h'^{-1},{\bf {\bar A}}\rightarrow h^{-1}\,({\bf {\bar A}}-\bar
\partial)\,h,
\ee where $ h=h(z,\bar z) $ and $ h'=h'(z,\bar z) $ are different group
elements in the null subgroup $ H=G_+ $ of $ G $.

The results obtained in the previous sections remain valid for the
case that the gauge groups are subgroups of $ G_+ $ and $ G_- $. Let us to
call these gauge groups $ H_+ $ and $ H_- $, with the constraint
$ H_-=\tau H_+ \tau^{-1} $, where $ \tau $ is the appropriate
involutive automorphism for $ G $ group that was constructed in the preceding
sections.
After
integrating out the gauge fields here the $ \sigma -$model analog of
the Toda theory interacting with the remaining directions of $ G_+ $ and $ G_-
$
that do not belong to $ H_+ $ and $ H_- $ respectively, is obtained
\cite{KLIMCIK}. 
Finally we would like to point out to an interesting connection between our
equivalence map and the well known duality map of symmetric Riemmanian
manifolds. Recall the way this duality maps appear in
Riemmanian geometry \cite{HELGASON}. For every Lie group $ G $ there are
involutive automorphisms of the Lie algebra $\cal {L}$$(G)$ which
have the additional property that convert the $\cal {L}$ $ (G_+) $ to
$\cal {L}$ $ (G_-) $. There are only
three different involutive
automorphisms for every simple algebra $\cal {L}$ $ (G) $ up to
conjugacy \cite{GILMORE}.

The complex conjugation operator $ C $, the operator $ I_{p,q} $ defined by
$\pmatrix {I_{p} & 0 \cr 0 & -I_{q} } $ and finally $ J_{p,p}=\pmatrix {0 &
I_{p}
\cr -I_{p} & 0 } \simeq \tilde J_{p,p}=\pmatrix {0 & \tilde I_{p} \cr - \tilde
I_{p} & 0 } $, where $ I_p  (\tilde I_p) $ is the $ p \times p $ unit
matrix
with $ +1 $ on the major (minor) diagonal and $ 0 $ elsewhere. These involutive
automorphisms was used to classify all real forms of the simple
algebras \cite{HELGASON}. In fact
let $ g $ be a compact simple algebra and $ \tau $ an involutive automorphism
of $ g $. Let $ g=f \oplus p $ be the cartan decomposition of $ g $ into
eigenspaces of the involutive automorphism $ \tau , \tau (f)=f , \tau (p)=-p $.
If we now
perform the Weyl unitary trick on the subspace $ p $, we get a new algebra $
g^* =f \oplus ip $, that is a real form of the complexification of $ g $ up to
isomorphism. The resulting Riemannian globally symmetric spaces $ G/F=EXP
(p) $ and $ G^*/F=EXP (ip) $ are said to be dual to each other
\cite{HERMANN}.

As an example take $ g $ to be $ su(2N,C) $ algebra
(the compact
real form of $ SL(2N,C) $). With
$ \tilde J_{N,N} C
$,
the maximal compact subalgebra $ f $ and the noncompact
algebra $ g^* $
are $ usp(2N) $ and $ su^*(2N) $, and the dual symmetric spaces
are $ SU(2N,C)/USP(2N) ,  SU^*(2N,C)/USP(2N) $. The operator $ \tilde J_{N,N} $
is in fact the same operator that was used in section three to map the
vector and chiral $ SL(2N,R)
$ \w to each other, $ \tau_{2N}=\tilde J_{N,N} $.

Similarly, for $ su(2 N+1,C) $, the dual automorphism is $
I_{N+1,N}=\tau_{2N+1} $ (up to conjugacy); for $ so(2N) $, it is $
J_{N,N}=s_{2N} $; and for $
so(2N+1) $, we have $ I_{1,2N}=(-)^{N} s_{2N+1} $ (up to conjugacy); and
finally for $ sp(2N,C) $ the corresponce is $ J_{N,N}=s_{2N} $.
\vskip .3in
\noindent{\bf{Acknowledgement}}
\vskip .1 in
The authors would like to thank Hessam Arfaei for helpfull discussions.
\vskip .1 in

\end{document}